\documentstyle[12pt,amsmath,amssymb,epic]{article}

\newcounter{resultcounter} 
\stepcounter{resultcounter}

\begin{document}

\title{\vspace*{-1cm}
Adiabatic Theorems \\ and Reversible Isothermal Processes}
\author{Walid K. Abou-Salem\footnote{supported by the Swiss National Foundation} and J\"urg Fr\"ohlich \\
Institute for Theoretical Physics \\
ETH-H\"onggerberg \\
CH-8093 Z\"urich, Switzerland}

\date{March $25^{th}$, 2005}
\maketitle

\begin{abstract}
Reversible isothermal processes of a finitely extended, driven quantum system
in contact with an infinite heat bath are studied from the point of view of
quantum statistical mechanics. Notions like heat flux, work and
entropy are defined for trajectories of states close to, but
distinct from states of joint thermal equilibrium. A theorem
characterizing {\it reversible} isothermal processes as {\it
quasi-static} processes (``{\it isothermal theorem}'') is
described. Corollaries concerning the changes of entropy and free
energy in reversible isothermal processes and on the $0^{th}$ law
of thermodynamics are outlined.
\end{abstract}

\vspace{0.5cm}

{\bf Mathematics Subject Classifications (2000).} 81Q05, 81Q10, 82C10
{\bf Key words.} adiabatic theorem, isothermal processes, reversible processes

\section{Introduction}
The problem to derive the fundamental laws of thermodynamics, the
$0^{th}$, $1^{st}$, $2^{nd}$ (and $3^{rd}$) law from kinetic
theory and non-equilibrium statistical mechanics has been studied
since the late $19^{th}$ century, with contributions by many
distinguished theoretical physicists including Maxwell, Boltzmann,
Gibbs and Einstein. In this note, we report on some recent results
concerning {\it isothermal processes} that have grown out of our
own modest attempts to bring the problem just described a little
closer to a satisfactory solution; (see [1] for a synopsis of our
results).

In the study of the $0^{th}$ law and of Carnot processes {\it
isothermal processes} play an important role. Isothermal processes, which are thermodynamic processes at constant temperature, arise
when a system with finitely many degrees of
freedom, the ``{\it small system}'', is coupled diathermally to an
infinitely extended, dispersive system, the ``{\it heat bath}'',
e.g., one consisting of black-body radiation, or of a gas of
electrons (metals), or of magnons (magnetic materials), at
positive temperature.\footnote{Diathermal contacts are local couplings that preserve all extensive quantities except the internal energy of the small subsystem. The latter will be precisely defined later.} During the past ten years, a particular
phenomenon, ``{\it return to equilibrium}'', encountered in the
study of isothermal processes of quantum-mechanical systems, has been analyzed for simple
(non-interacting) models of heat baths, by several groups of
people [2-6]: After coupling the small system to the heat bath,
the state of the coupled system approaches an equilibrium state at
the temperature of the heat bath, as time $t$ tends to $\infty$.
Further, if all contacts of the heat bath to its environment are
broken the state of the heat bath returns to its original
equilibrium state [7,8]. (The state of an isolated infinite heat bath is thus
characterized by a single quantity: its temperature!)

The results in [2-6] are proven using spectral- and resonance
theory, starting from the formalism developed in [9]. If
${\mathcal H}^{\mathcal R}$ denotes the Hilbert space of state
vectors of the infinite heat bath, ${\mathcal R}$, at {\it fixed}
temperature $(k\beta)^{-1}$, where $k$ denotes Boltzmann's
constant, (${\mathcal H}^{\mathcal R}$ is obtained with the help
of the GNS construction, see [9,10]), and ${\mathcal H}^\Sigma$
denotes the Hilbert space of {\it pure} state vectors of the ``small
system'' $\Sigma$, the Hilbert space of general (in particular
mixed) states of the {\it composed} system, ${\mathcal
R}\vee\Sigma$, is given by
\begin{equation}
{\mathcal H}\equiv {\mathcal H}^{{\mathcal R}\vee\Sigma}:={\mathcal H}^{\mathcal R}\otimes ({\mathcal H}^{\Sigma}\otimes {\mathcal H}^{\Sigma}) \; ;
\end{equation}
see [9,4]. The equilibrium state of the heat bath, or reservoir,
${\mathcal R}$ at inverse temperature $k\beta$ corresponds to a
vector $\Omega_\beta^{\mathcal R}\in {\mathcal H}^{\mathcal R}$,
and a general mixed state of the small system $\Sigma$ can be
described as the square root of a density matrix, which is a
Hilbert- Schmidt operator on ${\mathcal H}^{\Sigma}$, or, in other
terms, as a vector in ${\mathcal H}^{\Sigma}\otimes {\mathcal
H}^{\Sigma}$.

The dynamics of the composed system is generated by an (in
general, {\it time-dependent}) thermal Hamiltonian, or {\it
Liouvillian},
\begin{equation}
L(t):= L_0(t)+g(t)I \; ,
\end{equation}
(the ``standard Liouvillain'', see, e.g., [9,6]), where
\begin{equation}
L_0(t)=L_\beta^{\mathcal R}\otimes ({\mathbf 1}\otimes {\mathbf 1}) + {\mathbf 1}\otimes (H_0^{\Sigma}(t)\otimes {\mathbf 1})-{\mathbf 1}\otimes ({\mathbf 1}\otimes H_0^{\Sigma}(t))
\end{equation}
is the Liouvillian of the uncoupled system, $L_\beta^{\mathcal R}$
is the Liouvillian of the heat bath, with $L_\beta^{\mathcal
R}\Omega_\beta^{\mathcal R}=0$, $H_0^{\Sigma}(t)$ is the
(generally time-dependent) Hamiltonian of the small system, and
where $g(t)I$ is a spatially localized term describing the
interactions between ${\mathcal R}$ and $\Sigma$, with a
time-dependent coupling constant $g(t)$. Concrete models are
analyzed in [2-7].

We will only consider heat baths with a unique equilibrium state
at each temperature; (no phase coexistence). If $L_0(t)\equiv L_0$
and $g(t)\equiv g$ are independent of $t$, for $t\ge t_*$,
``return to equilibrium'' holds true if we can prove that $L$ has
a simple eigenvalue at $0$ and that the spectrum, $\sigma (L)$, of
L is purely continuous away from $0$; see [2,3,4]. The
eigenvector, $\Omega_\beta\equiv\Omega_\beta^{{\mathcal
R}\vee\Sigma}$, of $L$ corresponding to the eigenvalue $0$ is the
thermal equilibrium state of the coupled system ${\mathcal
R}\vee\Sigma$ at temperature $(k\beta)^{-1}$. Since $L_0$ tends to
have a rich spectrum of eigenvalues (embedded in continuous
spectrum), it is an a priori surprising consequence of
interactions between ${\mathcal R}$ and $\Sigma$ that the point
spectrum of $L$ consists of {\it only one simple eigenvalue} at
$0$; (see [2-6,11] for hypotheses on the interaction $I$ and
results). Under suitable hypotheses on ${\mathcal R}$ and
$\Sigma$, see [2,3,4], one can actually prove that ``return to
equilibrium'' is described by an {\it exponential law} involving a
finite {\it relaxation time}, $\tau_{\mathcal R}$.

If the Liouvillian $L(t)$ of ${\mathcal R}\vee\Sigma$ depends on
time $t$, but with the property that, for all times $t$, $L(t)$
has a simple eigenvalue at $0$ corresponding to an eigenvector
$\Omega_\beta(t)$, then $\Omega_\beta(t)$ can be viewed as an {\it
instantaneous equilibrium} (or {\it reference}) {\it state}, and
$\tau_{\mathcal R}(t)$ is called instantaneous relaxation time of
${\mathcal R}\vee\Sigma$. Let $\tau$ be the time scale over which
$L(t)$ changes appreciably. Assuming that, at some time $t_0$, the
state , $\Psi(t_0)$, of ${\mathcal R}\vee\Sigma$ is given by
$\Omega_\beta(t_0)$, it is natural to compare the state $\Psi(t)$
of ${\mathcal R}\vee\Sigma$ at a {\it later} time $t$ with the
instantaneous equilibrium state $\Omega_\beta(t)$ and to estimate
the norm of the difference $\Psi(t)-\Omega_\beta(t)$. One would
expect that if $\tau \gg \sup_{t} \tau_{\mathcal R}(t)$, then
\begin{equation*}
\Psi(t)\simeq \Omega_\beta (t) \; .
\end{equation*}

In [12], we prove an ``adiabatic theorem'', which we call ``{\it isothermal theorem}'', saying that
\begin{equation}
\Psi(t)\stackrel{\tau\rightarrow\infty}\rightarrow \Omega_\beta(t) \; ,
\end{equation}
for all times $t\ge t_0$. The purpose of our note is to carefully
state this theorem and various generalizations thereof and to
explain some of its consequences; e.g., to show that {\it
quasi-static} ($\tau\rightarrow\infty$) isothermal processes are
{\it reversible} and that, in the quasi-static limit, a variant of
the $0^{th}$ {\it law} holds. We also propose general definitions
of heat flux and of entropy for trajectories of states of
${\mathcal R}\vee\Sigma$ sampled in arbitrary isothermal processes
and use the ``isothermal theorem'' to relate these definitions to
more common ones. Details will appear in [1,12].

\section{A general ``adiabatic theorem''}

In this section we carefully state a general adiabatic theorem which is a slight improvement of results in [13,14] concerning adiabatic theorems for Hamiltonians without spectral gaps. Our simplest result follows from those in [13,14] merely by eliminating the superfluous hypothesis of semiboundedness of the generator of time evolution.

Let ${\mathcal H}$ be a separable Hilbert space, and let $\{ L(s)
\}_{s\in I}$, with $I\subset {\mathbb R}$ a compact interval, be a
family of selfadjoint operators on ${\mathcal H}$ with the
following properties:

(A1) The operators $L(s),s\in I,$ are selfadjoint on a common domain, ${\mathcal D}$, of definition dense in ${\mathcal H}$.

(A2) The resolvent $R(i,s):=(L(s)-i)^{-1}$ is bounded and differentiable, and $L(s)\dot{R}(i,s)$ is bounded uniformly in $s\in I$, where $\dot{( \; )}$ denotes the derivative with respect to $s$.

\vspace{0.5cm}

{\it Existence of time evolution.} If assumptions (A1) and (A2) hold then there exist unitary operators $\{ U(s,s')| s,s' \in I \}$ with the properties:

For all $s,s',s''$ in $I$,
\begin{equation*}
U(s,s)={\mathbf 1} \; , U(s,s')U(s',s'')=U(s,s'') \; ,
\end{equation*}
$U(s,s')$ is strongly continuous in $s$ and $s'$, and
\begin{equation}
i\frac{\partial}{\partial s}U(s,s')\Psi=L(s)U(s,s')\Psi \; ,
\end{equation}
for arbitrary $\Psi\in {\mathcal D}$, $s,s'$ in $I$; ($U$ is called a ``{\it propagator}'').

This result follows from, e.g., Theorems X.47a and X.70 in [15] in
a straightforward way; see also Theorem 2 of Chapter XIV in [16].
(Some sufficient conditions for (A1) and (A2) to hold are
discussed in [12].)

In order to prove an adiabatic theorem, one must require some
additional assumptions on the operators $L(s)$.

(A3) We assume that $L(s)$ has an eigenvalue $\lambda (s)$, that
$\{ P(s) \}$ is a family of finite rank projections such that
$L(s)P(s)=\lambda (s)P(s)$, $P(s)$ is twice continuously
differentiable in $s$ with bounded first and second derivatives,
for all $s\in I$, and that $P(s)$ is the spectral projection of
$L(s)$ corresponding to the eigenvalue $\lambda (s)$ for almost
all $s\in I$.

We consider a quantum system whose time evolution is generated by
a family of operators
\begin{equation}
L_\tau (t):= L(\frac{t}{\tau}) ,\; \frac{t}{\tau}=: s\in I \; ,
\end{equation}
where $\{ L(s) \}_{s\in I}$ satisfies assumptions (A1)-(A3). The
propagator of the system is denoted by $U_\tau(t,t')$. We define
\begin{equation}
U^{(\tau )}(s,s'):= U_\tau (\tau s, \tau s')
\end{equation}
and note that $U^{(\tau )}(s,s')$ solves the equation
\begin{equation}
i\frac{\partial}{\partial s}U^{(\tau)}(s,s')\Psi=\tau L(s) U^{(\tau)}(s,s')\Psi\; , \Psi\in {\mathcal D}\;.
\end{equation}

Next, we define
\begin{equation}
L_a(s):= L(s) +\frac{i}{\tau }[\dot{P}(s),P(s)]
\end{equation}
and the corresponding propagator, $U_a^{(\tau )}(s,s')$, which
solves the equation
\begin{equation}
i\frac{\partial }{\partial s} U_a^{(\tau )}(s,s')\Psi = \tau L_a(s)U_a^{(\tau )}(s,s')\Psi , \; \Psi\in {\mathcal D}\; .
\end{equation}
The propagator $U_a^{(\tau )}$ describes what one calls the {\it
adiabatic time evolution}. (Note that the operators $L_a(s), s\in
I$, satisfy (A1) and (A2), since, by (A3), $\frac{i}{\tau
}[\dot{P}(s),P(s)]$ are bounded, selfadjoint operators with
bounded derivative in $s$.)

\vspace{0.5cm}

{\it Adiabatic Theorem.} If assumptions (A1)-(A3) hold then
\begin{align*}
&\; \; \; \;(i) \; U_a^{(\tau )}(s',s)P(s)U_a^{(\tau )}(s,s')=P(s') \; (intertwining \; property)\; , \\
&{\mathrm for \; arbitrary \; } s,s'\; {\mathrm in \;} I , {\mathrm  \;and} \\
&\; \; \; \;(ii) \lim_{\tau\rightarrow\infty} \sup_{s,s'\in I} ||U^{(\tau )}(s,s')-U_a^{(\tau )}(s,s')||=0 \; .
\end{align*}

A proof of this result can be inferred from [14].

\vspace{0.5cm}

{\it Remarks.}

(1) We note that $U^{(\tau )}(s',s)=U^{(\tau )}(s,s')^*$.

(2) With more precise assumptions on the nature of the spectrum of
$\{ L(s) \}$, one can obtain information about the speed of
convergence in (ii), as $\tau \rightarrow \infty$; see [13,14,12].
A powerful strategy to obtain such information is to make use of
complex spectral deformation techniques, such as dilatation or
spectral-translation analyticity.

These techniques also enable one to prove an

(3) {\it Adiabatic Theorem for Resonances}, [17]. This result
resembles the adiabatic theorem described above, but eigenstates
of $L(s)$ are replaced by resonance states, and one must require
the adiabatic time scale $\tau$ to be small as compared to the
life time, $\tau_{res}(s)$, of a resonance of $L(s)$, uniformly in
$s\in I$. (For shape resonances, the techniques in [18] are
useful. Precise statements and proofs can be found in [17].)
Similar ideas lead to an adiabatic theorem in non-equilibrium
statistical mechanics, [22].

\section{The ``Isothermal Theorem''}

In this section, we turn to the study of isothermal processes of
``small'' driven quantum systems, $\Sigma$, in diathermal contact
with a heat bath, ${\mathcal R}$, at a fixed temperature
$(k\beta)^{-1}$. Our notations are as in Sect.1; see, eqs.(1),
(2), (3).

Let $L_\tau(t):=L(\frac{t}{\tau})$ denote the Liouvillian of the
coupled system ${\mathcal R}\vee\Sigma$, where $\{ L(s) \}_{s\in
I}$ is as in eqs.(2) and (3) of Sect.1 and satisfies assumptions
(A1) and (A2) of Sect.2. The interval
\begin{equation*}
I_{\tau }=\{ t \; |\;  \frac{t}{\tau }\in I\subset {\mathbb R} \}
\end{equation*}
is the time interval during which an isothermal process of
${\mathcal R}\vee\Sigma$ is studied.

We assume that $\Sigma$ is driven ``slowly'', i.e., that $\tau$ is
large as compared to the relaxation time $\tau_{\mathcal
R}=\max_{s\in I}\tau_{\mathcal R}(s)$ of ${\mathcal R}\vee\Sigma$.

Assumption (A3) of Sect.2 is supplemented with the following more specific assumption.

(A4) For all $s\in I\equiv [s_0,s_1]$, the operator $L(s)$ has a
{\it single, simple eigenvalue} $\lambda(s)=0$, the spectrum,
$\sigma(L(s))\backslash \{ 0 \}$, of $L(s)$ being {\it purely
continuous} away from 0. It is also assumed that, for $s\le s_0$,
$L(s)\equiv L$ is independent of $s$ and has spectral properties
sufficient to prove return to equilibrium [3,4].

Let $\Omega_\beta (s)\in {\mathcal H}$ denote the eigenvector of
$L(s)$ corresponding to the eigenvalue 0, for $s\le s_1$. Then
$\Omega_\beta(\frac{t}{\tau })$ is the {\it instantaneous
equilibrium state} of ${\mathcal R}\vee\Sigma$ at time $t$. Let
$P(s)= | \Omega_\beta (s)\rangle\langle \Omega_\beta(s) |$ denote
the orthogonal projection onto $\Omega_\beta (s)$; $P(s)$ is
assumed to satisfy (A3), Sect.2.

Let $\Psi(t)$ be the ``true'' state of ${\mathcal R}\vee\Sigma$ at time $t$; in particular
\begin{equation*}
\Psi(t)=U_\tau (t,t')\Psi(t')\; ,
\end{equation*}
where $U_\tau(t,t')$ is the propagator corresponding to $\{ L_\tau
(t) \}$; see eqs.(5)-(7), Sect.2. By the property of return to
equilibrium and assumption (A4),
\begin{equation}
\Psi(t)=\Omega_\beta\; , t\le \tau s_0 \; ,
\end{equation}
for an arbitrary initial condition $\Psi (-\infty )\in {\mathcal H}$ at $t=-\infty$.

We set
\begin{equation}
\Psi^{(\tau)}(s)=\Psi(\tau s) \; ,
\end{equation}
and note that, with the notations of eqs.(7), (8), Sect.2, and thanks to (11)
\begin{equation}
\Psi^{(\tau )}(s)=U^{(\tau)}(s,s_0)\Omega_\beta \; ,
\end{equation}
for $s\in I$.

\vspace{0.5cm}

{\it Isothermal Theorem.} Suppose that $L(s)$ and $P(s)$ satisfy
assumptions (A1)-(A3), Sect.2, and (A4) above. Then
\begin{equation*}
\lim_{\tau\rightarrow\infty}\sup_{s\in I} || \Psi^{(\tau )}(s) -\Omega_\beta(s) || = 0\; .
\end{equation*}

{\it Remarks.}

(1) The isothermal theorem follows readily from the adiabatic
theorem of Sect.2 and from eq.(13) and the definition of
$\Omega_\beta (s)$.

(2) We define the expectation values (states)
\begin{equation}
\omega_t^\beta (a):= \langle \Omega_\beta(\frac{t}{\tau }), a \Omega_\beta(\frac{t}{\tau }) \rangle
\end{equation}
and
\begin{equation}
\rho_t(a):= \langle \Psi(t), a \Psi(t)\rangle
\; ,
\end{equation}
where $a$ is an arbitrary bounded operator on ${\mathcal
H}={\mathcal H}^{{\mathcal R}\vee\Sigma}$. Then the isothermal
theorem says that
\begin{equation}
\rho_t(a)=\omega_t^\beta (a) + \epsilon_t^{(\tau )}(a)\; ,
\end{equation}
where
\begin{equation}
\lim_{\tau\rightarrow\infty}\frac{| \epsilon_t^{(\tau )} (a) | }{ || a ||} = 0 \; ,
\end{equation}
for all times $t\in I_\tau$.

(3) If the complex spectral deformation techniques of [2,3] are
applicable to the analysis of the coupled system ${\mathcal
R}\vee\Sigma$ then
\begin{equation}
|\epsilon_t^{(\tau )}(a)|\le {\mathcal O}(\tau^{-\frac{1}{2}}) ||a|| \; ;
\end{equation}
see [11,17].

(4) {\it All} our assumptions, (A1)-(A4), can be verified for the
classes of systems studied in [2-6] for which return to
equilibrium has been established therein. They can also be
verified for a ``quantum dot'' coupled to non-interacting
electrons in a metal, or for an impurity spin coupled to a
reservoir of non-interacting magnons; see [11,1,12].

\section{(Reversible) Isothermal Processes}

In this section, we study general isothermal processes and use the
isothermal theorem to characterize {\it reversible} isothermal
processes.

It will be convenient to view the heat bath ${\mathcal R}$ as the
{\it thermodynamic limit} of an increasing family of quantum
systems confined to compact subsets of physical space, as
discussed in [19,10]. Before passing to the thermodynamic limit of
the heat bath, the dynamics of the coupled system, ${\mathcal
R}\vee\Sigma$, is generated by a family of time-dependent
Hamiltonians
\begin{equation}
H(t)\equiv H^{{\mathcal R}\vee\Sigma}(t):= H^{\mathcal
R}+H^\Sigma(t)\; ,
\end{equation}
where
\begin{equation}
H^\Sigma (t)=H_0^\Sigma (t) + g(t)W\; ,
\end{equation}
$H_0^\Sigma (t)$ is as in Sect.1, and $W$ is the interaction
Hamiltonian (as opposed to the interaction Liouvillian, $I=ad_W$,
introduced in Sect.1).

Let ${\mathsf P}(t)$ denote the density matrix describing the
state of the coupled system, ${\mathcal R}\vee\Sigma$, at time
$t$, ({\it before} the thermodynamic limit for ${\mathcal R}$ is
taken). Then ${\mathsf P}(t)$ satisfies the Liouville equation
\begin{equation}
\dot{{\mathsf P}}(t)=-i[H(t),{\mathsf P}(t)] \; .
\end{equation}

The instantaneous equilibrium-, or reference state of the coupled
system is given, in the {\it canonical} ensemble, by the density
matrix
\begin{equation}
{\mathsf P}^\beta (t)=Z^\beta(t)^{-1}e^{-\beta H(t)} \; ,
\end{equation}
where
\begin{equation}
Z^\beta (t)= tr (e^{-\beta H(t)}) 
\end{equation}
is the partition function, and $tr$ denotes the trace. We assume that the thermodynamic limits
\begin{eqnarray}
\rho_t(\cdot)=TD\lim_{\mathcal R}tr ({\mathsf P}(t)(\cdot)) \\
\omega_t^\beta (\cdot) = TD\lim_{\mathcal R}tr ({\mathsf P}^\beta (t)(\cdot))
\end{eqnarray}
exist on a suitable kinematical algebra of operators describing ${\mathcal R}\vee\Sigma$; see [10,19,4].

The equilibrium state and partition function of a finitely extended heat bath are given by
\begin{eqnarray}
{\mathsf P}_{\mathcal R}^\beta= (Z_{\mathcal R}^\beta)^{-1}e^{-\beta H^{\mathcal R}}\; ,\\
Z_{\mathcal R}^\beta = tr(e^{-\beta H^{\mathcal R}}) \; ,
\end{eqnarray}
respectively.

Next, we introduce {\it thermodynamic potentials} for the small
system $\Sigma$: The {\it internal energy} of $\Sigma$ in the
``true'' state, $\rho_t$, of ${\mathcal R}\vee\Sigma$ at time $t$
is defined by
\begin{equation}
U^\Sigma (t):= \rho_t (H^\Sigma (t))
\end{equation}
and the {\it entropy} of $\Sigma$ in the state $\rho_t$ at time $t$ by
\begin{equation}
S^\Sigma (t):= -k \; TD\lim_{\mathcal R} tr({\mathsf P}(t)[ln {\mathsf P}(t)-ln{\mathsf P}_{\mathcal R}^\beta]) \; .
\end{equation}
Note that we here define $S^\Sigma (t)$ as a {\it relative}
entropy (with the aim of subtracting the divergent contribution of
the heat bath to the {\it total} entropy). A general inequality
for traces, see [10], says that
\begin{equation}
S^\Sigma (t)\le 0 \; .
\end{equation}

The {\it free energy} of $\Sigma$ in an {\it instantaneous
equilibrium state}, $\omega^\beta_t$, of ${\mathcal R}\vee\Sigma$
is defined by
\begin{equation}
F^\Sigma (t):= -kT \; TD\lim_{\mathcal R}ln (\frac{Z^\beta (t)}{Z^\beta_{\mathcal R}}) \; .
\end{equation}

Next, we define quantities associated not with states but with the
{\it thermodynamic process} carried out by ${\mathcal
R}\vee\Sigma$: the {\it heat flux} into $\Sigma$ and the {\it work
rate}, or {\it power}, of $\Sigma$. Let $\delta$ denote the so
called ``imperfect differential''. Then
\begin{equation}
\frac{\delta Q^\Sigma}{dt}(t):= TD\lim_{\mathcal R}-\frac{d}{dt}tr ({\mathsf P}(t)H^{\mathcal R})\; ,
\end{equation}
and
\begin{equation}
\frac{\delta A^\Sigma}{dt}(t):= \rho_t(\dot{H}^\Sigma (t)) \; ;
\end{equation}
see [12,1,8] for details.

We are now prepared to summarize our main results on isothermal processes. The first two results, which apply to general isothermal processes, are related to the first law of thermodynamics and the relationship between the rate of change of entropy production and the heat flux into $\Sigma$. The remaining three results are corollaries of the isothermal theorem pertaining to free energy and the change of entropy in reversible isothermal processes and the zeroth law of thermodynamics.

(1) From definitions (28), (32) and (33) and the Liouville equation (21) it follows that
\begin{equation}
\dot{U}^\Sigma (t)=\frac{\delta Q^\Sigma}{dt}(t)+\frac{\delta A^\Sigma}{dt}(t)\; ,
\end{equation}
which is the {\it first law of thermodynamics}; (hardly more than a definition of $\frac{\delta A^\Sigma}{dt}$).

(2) Note that, by the unitarity of time evolution and the cyclic invariance of the trace,
\begin{equation*}
\frac{d}{dt}tr ({\mathsf P}(t)ln {\mathsf P}(t))=0 \; ,
\end{equation*}
and
\begin{equation*}
\frac{d}{dt} tr ({\mathsf P}(t)ln Z^\beta_{\mathcal R})=\frac{d}{dt}lnZ^\beta_{\mathcal R}=0\; .
\end{equation*}
Together with definitions (26), (29) and (32), this implies that
\begin{equation}
\dot{S}^\Sigma (t)=\frac{1}{T} \frac{\delta Q^\Sigma}{dt} (t)\; ,
\end{equation}
for {\it arbitrary} isothermal processes at temperature $T=(k\beta)^{-1}$.

(3) Next, we consider an isothermal process of ${\mathcal
R}\vee\Sigma$ during a finite time interval $I_\tau = [\tau s_0,
\tau s_1]$, with $s_0$ and $s_1$ fixed. The initial state
$\rho_{\tau s_0}$ of ${\mathcal R}\vee\Sigma$ is assumed to be an
equilibrium state $\omega^\beta_{\tau s_0}$ of the Liouvillian
$L_\tau (\tau s_0)=L(s_0)$. We are interested in the properties of
such a process when $\tau$ becomes large, i.e., when the process
is {\it quasi-static}.

\vspace{0.5cm}

{\it Result.} Quasi-static isothermal processes are {\it
reversible} (in the sense that all intermediate states $\rho_t$ of
${\mathcal R}\vee\Sigma$, $t\in I_\tau$, converge in norm to {\it
instantaneous equilibrium states} $\omega^\beta_t$, as
$\tau\rightarrow\infty$).

This result is an immediate consequence of the isothermal theorem.
It means that, for all practical purposes, an isothermal process
with time scale $\tau$ is reversible if $\tau \gg \tau_{\mathcal
R} = \max_{s\in I}\tau_{\mathcal R}(s)$.

(4) For reversible isothermal processes, the usual {\it
equilibrium definitions} of internal energy and entropy of the
small system $\Sigma$ can be used:
\begin{equation}
U_{rev}^\Sigma (t) := \omega^\beta_t (H^\Sigma (t))\; ,
\end{equation}
\begin{equation}
S_{rev}^\Sigma (t):= -k \; TD\lim_{\mathcal R}tr ({\mathsf P}^\beta (t)[ln {\mathsf P}^\beta (t)-ln {\mathsf P}^{\mathcal R}])=\frac{1}{T}(U_{rev}^\Sigma (t)-F^\Sigma (t))\; ,
\end{equation}
where the free energy $F^\Sigma (t)$ has been defined in (31), and the second equation in (37) follows from (22), (26), (31) and (36). Eqs.(37) and (31) then imply that
\begin{equation*}
\dot{S}^\Sigma_{rev}(t)=\frac{1}{T} (\frac{d}{dt}\omega^\beta_t (H^\Sigma (t))-\omega^\beta_t (\dot{H}^\Sigma (t)))\; .
\end{equation*}
Recalling (34) and (35), and applying the isothermal theorem, we find that
\begin{align}
\dot{S}^\Sigma (t) &\rightarrow \dot{S}^\Sigma_{rev}(t) \; , \\
\frac{\delta A^\Sigma}{dt}(t) &\rightarrow \dot{F}^\Sigma (t) \; ,
\end{align}
as $\tau\rightarrow\infty$.

(5) We conclude this overview by considering a quasi-static
isothermal process of ${\mathcal R}\vee\Sigma$ with $H^\Sigma
(s)\rightarrow H^\Sigma_0, \; g(s)\rightarrow 0$, as $s\nearrow
s_1$, i.e., the interactions between ${\mathcal R}$ and $\Sigma$
are switched off at the end of the process. Then the isothermal
theorem implies that
\begin{equation}
\lim_{\tau\rightarrow\infty}\lim_{s\nearrow s_1}\rho_{\tau s} = \omega^\beta_{\mathcal R}\otimes \omega^\beta_\Sigma \; ,
\end{equation}
where $\omega^\beta_{\mathcal R}(\cdot)=(Z^\beta_{\mathcal R})^{-1}tr (e^{-\beta H^{\mathcal R}}\cdot )$, see (26), and
\begin{equation}
\omega^\beta_\Sigma(\cdot )=(Z^\beta_\Sigma)^{-1}tr (e^{-\beta H_0^\Sigma} \cdot )
\end{equation}
is the Gibbs state of the small system $\Sigma$ at the temperature
$(k\beta)^{-1}$ of the heat bath, {\it independently} of the
properties of the diathermal contact (i.e., of the interaction
Hamiltonian $W$), assuming that (A1)-(A4) hold for $s<s_1$.

This result and the property of return to equilibrium for the heat
bath ${\mathcal R}$ yield, in essence, the $0^{th}$ law of
thermodynamics.

Carnot processes and the $2^{nd}$ law of thermodynamics are
discussed in [1]; (see also [10,20,21]). An important variant of
the adiabatic theorem for non-equilibrium stationary states will
appear in [22]. The analysis in [22] is based on some basic
techniques developed in [23].

\vspace{1cm}

{\bf References}
\vspace{0.5cm}

\noindent
1. Abou-Salem, W.K. and Fr\"ohlich, J.: {\em Status of the fundamental laws of thermodynamics}, in preparation.

\noindent
2. Jaksi\'c, V. and Pillet, C.-A.: {\em On a model of quantum friction II: Fermi's golden rule and dynamics at positive temperature}, Comm. Math. Phys. {\bf 176} (1996), 619-644. 

\noindent
3. Jaksi\'c, V. and Pillet, C.-A.: {\em On a model of quantum friction III: Ergodic properties of the spin-boson system}, Comm. Math. Phys. {\bf 178} (1996), 627-651.

\noindent
4. Bach, V., Fr\"ohlich, J. and Sigal, I.M.: {\em Return to equilibrium}, J. Math. Phys. {\bf 41} (2000), 3985-4060. 

\noindent
5. Merkli, M.: {\em Positive commutators in non-equilibrium quantum statistical mechanics}, Comm. Math. Phys. {\bf 223} (2001), 327-362. 

\noindent
6. Fr\"ohlich, J. and Merkli, M.: {\em Another return of ``return to equilibrium''}, Comm. Math. Phys. {\bf 251} (2004), 235-262. 

\noindent
7. Robinson, D.W.: {\em Return to equilibrium}, Comm. Math. Phys. {\bf 31} (1973), 171-189. 

\noindent
8. Fr\"ohlich, J., Merkli, M., Ueltschi, D. and Schwarz, S.: {\em Statistical mechanics of thermodynamic processes}, in {\em A garden of quanta}, 345-363, World Sci. Publishing, River Edge, New Jersey, 2003.

\noindent
9. Haag, R., Hugenholtz, N.M. and Winnink, M. : {\em On equilibrium states in quantum statistical mechanics}, Comm. Math. Phys. {\bf 5} (1967), 215-236. 

\noindent
10. Bratelli, O. and Robinson, D.W.: {\em Operator algebras and quantum statistical mechanics I,II}, Texts and Monographs in Physics, Springer-Verlag, Berlin, 1987.

\noindent
11. Abou-Salem, W.K.: {\em PhD thesis} (2005).

\noindent
12. Abou-Salem, W.K. and Fr\"ohlich, J., in preparation.

\noindent
13. Avron, J.E. and Elgart, A.: {\em Adiabatic theorem without a gap condition}, Comm. Math. Phys. {\bf 203} (1999), 445-463. 

\noindent
14. Teufel, S.: {\em A note on the adiabatic theorem}, Lett. Math. Phys. {\bf 58} (2001), 261-266. 

\noindent
15. Reed, M. and Simon, B.: {\em Methods of modern mathematical physics}, vol.II, Academic Press, New York, 1975.

\noindent
16. Yosida, K.: {\em Functional analysis}, 6th ed., Springer-Verlag, Berlin, 1998.

\noindent
17. Abou-Salem, W.K. and Fr\"ohlich, J.: {\em Adiabatic theorem for resonances}, in preparation.

\noindent
18. Fr\"ohlich, J. and Pfeifer, P.: {\em Generalized time-energy uncertainty relations and bounds on lifetimes of resonances}, Rev. Mod. Phys. {\bf 67} (1995), 759-779.

\noindent
19. Ruelle, D.: {\em Statistical mechanics: rigorous results}, World Scientific, Singapore, 1999.

\noindent
20. Ruelle, D.: {\em Entropy production in quantum spin systems}, Comm. Math. Phys. {\bf 224} (2001), 3-16. 

\noindent
21. Fr\"ohlich, J., Merkli, M. and Ueltschi, D.: {\em Dissipative transport: thermal contacts and tunnelling junctions}, Ann. Henri Poincar\'e {\bf 4} (2003), 897-945. 

\noindent
22. Abou-Salem, W.K.: {\em An adiabatic theorem for non-equilibrium steady states}, in preparation.

\noindent
23. Jaksi\'c, V. and Pillet, C.-A.:{\em Non-equilibrium steady states of finite quantum systems coupled to thermal reservoirs}, Comm. Math. Phys. {\bf 226} (2002), 131-162.

\end{document}